\documentclass[11pt]{report}
\usepackage{psfig}
\title{The Dynamic Aperture and the High Multipole Limit}
\author{G. Parzen}
\date{February 6, 1998 \\BNL-65364}
\begin{document}
\maketitle
\tableofcontents
\begin{abstract}
\noindent Tracking studies have indicated  that for a lattice whose elements 
all have  a single field multipole present, all having the same order $k$,
the dynamic aperture  approaches a  non zero limit when $k$  becomes very 
large. The dynamic aperture and other properties of the   lattice, as $k$ 
becomes large, will be called the high multipole limit. It will be shown that 
the high multipole limit  provides a reasonable estimate of the dynamic 
aperture of an accelerator, and the other properties of the high multipole 
limit found below are useful for understanding the stability of the 
accelerator. The high multipole limit is  easily computed and it also 
provides an estimate of  how  much can be gained by correcting the lower 
field multipoles.  The above results will be illustrated by tracking studies 
done with a simple one cell lattice, and with a RHIC lattice having six low 
beta insertions.
\end{abstract}

%\section{Introduction}
\pagestyle{headings}
\chapter{Introduction}

Tracking studies have indicated \cite{key1} that for a lattice whose elements 
all have a single field multipole present, all having the same order $k$, the 
dynamic aperture  approaches a  non-zero limit when $k$  becomes very 
large. The dynamic aperture and other properties of the lattice, as $k$ 
becomes large,  will be called the high multipole limit. It will be shown 
that the high multipole limit  provides a reasonable estimate of the dynamic 
aperture of an accelerator, and the other properties of the high multipole 
limit found below are useful for understanding the stability of the 
accelerator. The high multipole limit is  easily computed and it also 
provides an estimate of  how  much can be gained by correcting the lower 
field multipoles.

The above results will be illustrated by tracking studies done with a 
simple one cell lattice, and with a RHIC lattice having six low beta 
insertions. The properties of the high multipole limit  that are 
demonstrated in these tracking studies can be used to answer the following 
kinds of questions about the dynamic aperture:
\begin{enumerate}
\item
Up to which order multipole does one have to correct  to regain the 
aperture loss due to the field multipoles present.
\item
How much is the aperture loss due to the field multipoles present.
\end{enumerate}

The following are some properties of the high multipole limit  discussed 
below which can be useful in understanding the stability of a given lattice:
\begin{enumerate}
\item
The stability boundary in the high multipole limit is dominated and 
determined by one set of elements in the lattice,  and that is the set of 
elements which have the smallest value of $R/\beta^{0.5}$, where $\beta$ is 
the linear beta function at the element and $R$ is the multipole parameter 
that corresponds to the magnet radius.
\item
Analytical results can be found for the stability boundary in the high 
multipole limit which can be used to estimate the loss in aperture due to 
the multipoles present in the lattice.
\item
The stability boundary in the high multipole limit  does not depend on the 
strength of the multipoles or on the choice of the linear tunes, $\nu_{x}$ 
and $\nu_{y}$. This leads to the suggestions that the stability boundary 
of a  lattice is insensitive to the magnitude of the higher multipoles, 
and the linear tune needs to be chosen to avoid the resonances driven by 
the lower order multipoles. Higher and lower multipoles are defined below.
\end{enumerate}

\chapter{The high multipole limit in  2-dimensions}

This section will be devoted to establishing the basic rule for the high 
multipole limit in 2-dimensional phase space.  Consider a linear periodic 
lattice where each element of the lattice is perturbed by a single non-linear 
field multipole, and the order of this multipole, $k$, is the same in each 
element.  The field multipole will produce a field in each element 
whose vertical component, $B_y$, in the median plane is given by
\begin{equation}
B_y=B_0\ b_k\ x^{k}, \qquad b_k=\mathrm{b}/R^{k}
\label{eq:first}
\end{equation}
In computing the high multipole limit, we will be computing the dynamic
aperture of this lattice for different values of $k$, and in particular for
large values of $k$. 
$R$ may vary from element to element, but does not vary with $k$. $\mathrm{b}$ may vary 
from element to element,  and $\mathrm{b}$ may also vary with $k$ but not by large 
factors.The dominant variation in $b_k$ with $k$ is given by the $1/R^{k}$  
factor.(More exactly, $(\Delta \mathrm{b}/\mathrm{b})^{1/k}$  approaches \ 1 for large enough
$k$, where  $\Delta \mathrm{b}$ is the largest change in $\mathrm{b}$  either from element to 
element or as a function of $k$.)
 For an actual accelerator, R may be chosen as the 
radius of the magnet coil, and the measured $\mathrm{b}=b_k R^{k}$ has to satisfy the 
above conditions in order to apply the high multipole limit results found below. 
Let $x_0$, $p_{x0}$ be the initial particle coordinates. One may define the 
stability boundary to be a closed curve in $x_0$, $p_{x0}$ such that for any 
choice of $x_0$, $p_{x0}$ outside this boundary the particle motion for a given
number of periods will be considered unstable. The definition of stability 
is discussed in section 6.  The basic rule may now be stated as follows:

\bigskip

\noindent\underline{Basic rule for the high multipole limit in 2 dimensional 
phase space.}

\medskip

For a particle moving through a linear periodic lattice in the 2 dimensional 
phase space of $x$, $p_x$, in the presence of non linear field multipoles $b_k$, 
which have the form
\begin{equation}
 b_k=\mathrm{b}/R^{k}
\label{eq:second}
\end{equation}
the stability boundary which encloses the stable area in $x_0$, $p_{x0}$, for 
a given number of periods and for large enough $k$, is given by
\begin{equation}
\epsilon(x_0,p_{x0})=[R^2/\beta_x]_{\rm{min}}
\label{eq:third}
\end{equation}
\[
\epsilon(x,p_x)=\gamma_x \ x^2+2 \alpha_x \ x \ p_x +\beta_x \ p_x^2
\]
$[R^2/\beta_x]_{\rm{min}}$ is the minimum value of $R^2/\beta_x$ in the 
elements of the lattice where $b_k$ is not zero, and $\gamma_x$, $\alpha_x$, 
$\beta x$ are the linear parameters of the lattice.

An analytical argument can be given which indicates what lies behind 
Eq.~\ref{eq:third}.  For very large $k$, the multipole  field as given by 
Eq.~\ref{eq:first}  
approaches zero when $x$ in any element is smaller than the $R$ value of 
that element. Thus for small enough $x, \epsilon(x,p_x)$ becomes a constant 
of the motion. Some thought will then show that the element with the 
smallest value of $R^2/\beta_x$ determines the largest emittance that is 
stable, which is given by Eq.~\ref{eq:third}. A possible flaw in this argument is 
that for a given $k$, no matter how large, when $x$ gets close enough 
to the stability boundary , the multipole field can become appreciably 
different from zero.

The above basic rule will be justified below by doing a number of numerical 
tracking experiments, tracking particles through a number of different 
lattices.  Most of the tracking experiments reported below are done with 
a simple one cell lattice. The results found will be further illustrated 
with results for a RHIC lattice with 6 low beta insertions.

%\newpage
\bigskip

\noindent\underline{The simple one cell lattice}

\medskip

The simple one cell lattice initially used in this study consists of a 
focussing quadrupole, $q_f$, and a defocussing quadrupole, $q_d$, separated 
by drift spaces of equal length.  The perturbing non-linear field multipole 
is initially placed in the middle of $q_f$. The observation  point for 
measuring the dynamic aperture is initially chosen to be at the middle of 
the perturbing field multipole. This simple one cell lattice will be 
referred to as the simple one cell lattice.

The perturbing field multipole is a point multipole which produces a 
vertical field on the median plane whose integrated strength, field times 
length, is given by
\begin{equation}
B_0 \ \mathrm{b} \ x^k / R^k
\label{eq:fourth}
\end{equation}
The parameters $B_0$, $\mathrm{b}$, $R$ are usually chosen to make the lattice 
resemble the RHIC lattice, with nonlinear effects of the same order as 
those seen in RHIC. To examine  the high multipole limit one will be 
particularly interested in finding the dynamic aperture for large values 
of $k$.

The transfer functions, that give the final particle coordinates for a given 
set of its initial coordinates for each element in the lattice, are given in 
section 7. For the reasons given there, the exact equations of motion are 
used in finding the transfer functions. The parameters of the quadrupole 
and drift spaces are initially chosen to produce the tune $\nu_{x0}=0.1740$. 
This tune was chosen to lie in a region free of all resonances up to the 
tenth order, and it lies between the 1/5 and 1/7 resonances. The lattice 
parameters are given in section 7.

\bigskip

\noindent\underline{$x_{sl0}$ vs $k$ at $q_f$ for the simple lattice with a 
single $b_k$ at $q_f$}

\medskip

\begin{figure}[tbh]
\centerline{\psfig{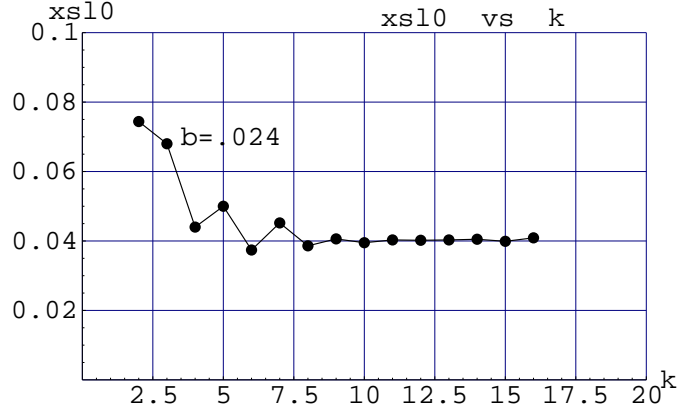}}
\caption{A plot of $x_{sl0}$ versus $k$ for the simple one cell lattice with 
a single  multipole of order $k$ at $q_f$.  $x_{sl0}$ is the largest $x_0$ that 
is stable for 100 periods.  $p_{x0}=p_{y0}=y_0=0$ and 2 dimensional motion.  
In the figure xs10 represents  $x_{sl0}$ .}
\label{fig:figone}
\end{figure}

The basic rule for the hml (high multipole limit), and Eq.~\ref{eq:third} 
will be illustrated by doing tracking runs using the simple one cell lattice. 
A single field multipole $b_k$, as given by Eq.~\ref{eq:fourth}, will be 
placed in the middle of the focusing quadrupole $q_f$. The particle will 
be started with $p_{x0}=0$, and $x_0$ will be varied to find the largest $x0$ 
that is stable for 100 periods, which will be denoted by $x_{sl0}$. This will 
be done for different values of $k$, the order of the multipole.  According 
to Eq.~\ref{eq:third}, since in this case $p_{x0}=0$, $\alpha_x=0$, one should find 
for large enough $k$ that $x_{sl0}=R$. Here $R$  was chosen as $R=0.04$ m. 
The results are shown in Fig.~\ref{fig:figone}.

One sees that at lower values of $k$, $x_{sl0}$ varies rapidly and then levels 
out at a value close to $x_{sl0}=R=0.04$. The high multipole limit gives a good 
approximation for $x_{sl0}$ starting with the relatively low values of $k>10$ 
for non-linear multipoles of the order of those expected in RHIC.  It will 
be seen below that this property helps to make the high multipole limit 
useful for estimating the dynamic aperture of an accelerator.

\bigskip

\noindent\underline{The stability surface at $q_f$ for the simple lattice 
with a single $b_k$ at $q_f$}

\medskip

The stability boundary in the high multipole limit will now be found for 
the simple one cell lattice with just one $b_k$ at $q_f$. To find the 
stability surface in $x_0$, $p_{x0}$, one can search along different directions 
in $x_0,p_{x0}$ space to find the stability boundary in that direction.  One 
may write the initial coordinates as
\begin{eqnarray}
x_0  & = & (\beta_x\ \epsilon_{x0})^{0.5} cos(\alpha \pi/2) \nonumber \\
p_{x0} & = & (\epsilon_{x0} /\beta_x)^{0.5} sin(\alpha\pi/2) 
\label{eq:fifth}
\end{eqnarray}
$\alpha$ gives the direction of search in $x_0$, $p_{x0}$ and $\epsilon_{x0}$ is 
the initial linear emittance for this choice of $x_0$. $p_{x0}$ at $q_f$ where 
$\alpha_x=0$, $\beta_x=68.4497$ m. To do the search $\epsilon_{x0}$ will be 
increased  until the motion becomes unstable for a particular choice of 
$\alpha$. The value of $\epsilon_{x0}$ that lies on the stability boundary 
for this search direction will be denoted by $\epsilon_{xsl0}$. According to 
the basic rule for the high multipole limit, Eq.~\ref{eq:third}, we should 
find that $\epsilon_{xsl0}$ is constant for all directions at the value 
$\epsilon_{xsl0}=R^2/\beta_x=23.37482740$. This would establish the basic 
rule for this lattice.  This tracking study is done with a single field 
multipole, $b_k$, at the middle of $q_f$.  In order to find the stability 
boundary for the high multipole limit, $k$ is chosen at the large value 
of $k=1\cdot 10^6$. The results of this tracking study are shown in  
Table~\ref{tab1}, where $\epsilon_{xsl0}$ is shown as a function of $\alpha$.

\begin{table}[tbh]
\begin{center}
\begin{tabular}{|c|c|}\hline
$\alpha$ & $\epsilon_{xsl0}$ \\ \hline
$-1$	&	23.3893 \\
$-0.75$	&	23.3749 \\
$-0.5$	&	23.3756 \\
$-0.25$	&	23.3748 \\
$0.$	&	23.3747 \\
$0.25$	&	23.386 \\
$0.5$	&	23.4199 \\
$0.75$	&	23.4024 \\
$1.$	&	23.3893 \\ \hline
\end{tabular}
\end{center}
\caption{The linear emittance, $\epsilon_{xsl0}$, computed for different 
points, $x_0$, $p_{x0}$ on the stability boundary in the hml, corresponding to 
different directions in $x_0$, $p_{x0}$ space, at the location of $q_f$ in 
the simple one cell lattice.  The parameter $\alpha$ gives the different 
directions according to Eqs.~\ref{eq:fifth}.  $\epsilon_{xsl0}$ is in 
mm mrad.}
\label{tab1}
\end{table}

One sees in Table~\ref{tab1} that $\epsilon_{xsl0}$ for different directions
is almost constant at the value 23.37482740 as predicted by the basic
rule.  However, there is a variation in $\epsilon_{xsl0}$ of about .2\%,
which appears to show that the basic rule for high multipole limit,
Eq.~\ref{eq:third}, is not exact but has a small error in it.  This
error is not important for the main results of this paper.  This
particular study shows that it is convenient to have a precise
definition of stability such as is given in section 6.

\bigskip

\noindent\underline{$x_{slo}$ at $q_d$ for the simple lattice with a single $b_k$
at $q_f$}

\medskip
Our next step will be to consider the stability limit at some other location 
in the lattice, where $\beta_x$ is not at its maximum value, such as at $q_d$ 
where $\beta_x$ is $\beta_x=21.1234 5678$. According to the basic rule for 
the high multipole limit for the simple one cell lattice, the stability limit 
in $x_0$ when $p_{x0}=0$ at $q_f$, $x_{sl0f}$, is related to the stability limit at 
$q_d$, $x_{sld}$, at large enough $k$, by
\begin{equation}
x_{sl0d}= (\beta_{xd}/\beta_{xf})^{0.5}\   x_{sl0f}
\label{eq:sixth}
\end{equation}
where $\beta_{xf}$ and $\beta_{xd}$ are the beta functions at $q_f$ and $q_d$. 
This is illustrated in Table~\ref{tab2} where $x_{slf}$ and $x_{sld}$ are 
compared as a function of $k$, the order of the field multipole at $q_f$.
\begin{table}[tbh]
\begin{center}
\begin{tabular}{|c|c|c|c|}\hline
$k$	&	$x_{slqd}$	&	$x_{slqf}$	&	$[x_{slqd}/x_{slqf}] \sqrt{\beta _{xf}/\beta_{xd}}$ \\ \hline
1000000		& 0.02248406	& 0.03999994	& 1.0000187 \\
10000000	& 0.0224841	& 0.03999999	& 1.0000193 \\
100000000	& 0.02248411	& 0.03999999	& 1.0000198 \\ \hline
\end{tabular}
\end{center}
\caption{A table showing how well $x_{sl0}$ in the hml varies like 
$\beta_x^{0.5}$ through the lattice. $x_{slo}$ is the largest $x_0$ that is 
stable for 100 periods for the simple one cell lattice  when $p_{x0}=0$ with 
a single multipole of order $k$ at $q_f$.  $x_{sl0}$ at $q_f$ and $q_d$ are 
indicated by $x_{sl0f}$ and $x_{sl0d}$, and $\beta_{xqf}$ and $\beta_{xqd}$ are the 
linear beta functions at $q_f$ and $q_d$.  All lengths are in meters.}
\label{tab2}
\end{table}

\noindent One sees from Table~\ref{tab2} that the prediction of the basic 
rule for this case, as given by Eq.~\ref{eq:sixth}, is valid with an error 
of about $2\ 10^{-5}$. The reason for varying $k$ was to show that the error 
found was not due to $k$ not being large enough as the basic rule for the 
high multipole limit holds only for large enough $k$ values. The error in 
computing the beta functions also appears to be too small to account for 
the error found in Table~\ref{tab2}.

\bigskip

\noindent\underline{$x_{slo}$ at $q_d$ and $q_f$ for the simple lattice with 
a single $b_k$ at $q_f$ and a $b_k$ at} \\
\underline{$q_d$ with a different $R$ value}

\medskip

We will now consider the case of the simple one cell lattice with two 
field multipoles present; one at $q_f$ and one at $q_d$. The field multipole 
at $q_d$ will have an $R$-value (see Eq.~\ref{eq:fourth}), $R_{qd}$ which is 
different from the $R$-value, $R_{qf}$, of the multipole at $q_f$.  According 
to the basic rule for the hml, Eq.~\ref{eq:third}, when $R_{qd}=R_{qf}$, the 
multipole at $qf$ will dominate in determining the hml since 
$[R^2/\beta_x]_{\rm min}$ will occur at $q_f$ where $\beta_x$ has its 
maximum.  In particular, Eq.~\ref{eq:third} gives $x_{sl0}$ when $p_{s0}=0$ 
as $x_{sl0f}=R_{qf}$, 
while $x_{sl0}$ at $q_d$ is given by Eq.~\ref{eq:sixth} as $x_{sl0d}=
\sqrt{[\beta_{xd}/\beta_{xf}]} Rqf$.  If one now reduces $R_{qd}$, the 
multipole at $q_f$ will continue to dominate until $R_{qd}$ reaches the
 value $\sqrt{[\beta_{xd}/\beta_{xf}]} R_{qf}$,  and then the multipole at 
$q_d$ will start to dominate, and $x_{sl0d}$  will be given by $R_{qd}$ 
while $x_{sl0f}$ will be given by $\sqrt{[\beta_{xf}/\beta_{xd}]} R_{qd}$.  
These results are illustrated by the computed results given in 
Table~\ref{tab3} where $x_{sl0f}$ and $x_{sl0d}$ are shown as 
$R_{qd}/R_{qf}$  is decreased from 1 to 0.2. For this lattice $R_{qf}$ is 
held constant at 0.04 m and $\sqrt{[\beta_{xd}/\beta_{xf}]}=0.56209177$.

\begin{table}[tbh]
\begin{center}
\begin{tabular}{|c|c|c|c|}\hline
$R_{qd}/R_{qf}$	& $x_{sl0qf}$	& $R_{qd}$		& $x_{sl0qd}$ \\ \hline
1.		& 0.03999	& 0.04		& 0.02248 \\
0.9		& 0.03999	& 0.036		& 0.02248 \\
0.8		& 0.03999	& 0.032		& 0.02248 \\
0.7		& 0.03999	& 0.028		& 0.02248 \\
0.6		& 0.03999	& 0.024		& 0.02248 \\
0.55		& 0.03914	& 0.022		& 0.02199 \\
0.5		& 0.03558	& 0.02		& 0.01999 \\
0.4		& 0.02846	& 0.016		& 0.01599 \\
0.3		& 0.02134	& 0.012		& 0.01199 \\ \hline
\end{tabular}
\end{center}
\caption{A table showing that the stability boundary in the hml is 
determined by the element in the lattice with the smallest value of 
$R/\beta_x^{0.5}$. Results shown are for the simple one cell lattice with a 
single multipole with the same order at $q_f$ and $q_d$. $R_{qf}$, $R_{qd}$ and 
$x_{sl0f}$, $x_{sl0d}$ are the $R$ and $x_{sl0}$ parameters at $q_f$ and 
$q_d$.  All lengths are in meters.}
\label{tab3}
\end{table}

It will be suggested below that the high multipole limit provides a 
reasonable measure of the dynamic aperture.  Assuming this to be so,  
then the basic rule for the hml states that in an accelerator the dynamic 
aperture is dominated by the magnet with the smallest $R/\beta_x^{0.5}$.  
For example, if an accelerator has 6 insertions with 6 crossing points, 
with similar magnets that are excited to give different $\beta_x$ at the 
crossing points, the magnet which has the largest $\beta_x$ will dominate 
and determine the dynamic aperture\cite{key2}.  This magnet is located 
in the insertion with the smallest $\beta_x$ at the crossing point.  
The dynamic aperture at other locations in the lattice will scale like 
$(\beta_x/\beta_{x{\rm max}})^{0.5}$ where $\beta_{x{\rm max}}$ 
is the largest $\beta_x$ in the lattice.

\bigskip

\noindent\underline{The dependence of $x_{sl0}$ vs $k$ at $q_f$ on the 
strength of field multipole, $\mathrm{b}$}

\medskip

In the previous tracking studies, the multipole present is assumed to have 
the form, $b_k=\mathrm{b}/R^k$, where $\mathrm{b}$ is the integrated strength of the point 
multipole.  $\mathrm{b}$ and $R$ can vary from one element in the lattice to another. 
The basic rule for the high multipole limit, Eq.~\ref{eq:third}, does not 
show any dependence on $\mathrm{b}$.  Tracking studies show that when $\mathrm{b}$ is 
increased, the high multipole limit is unchanged, but one must go to a 
larger value of $k$ before the high multipole limit is reached. The studies 
also show that for large enough values of $k$, $x_{sl0}$, the largest stable 
$x_0$ when $p_{x0}=0$, goes like $\mathrm{b}^{(1/k)}$.  This indicates that for large 
enough $k$, $x_{sl0}$ is insensitive to the size of $\mathrm{b}$.  For RHIC, the high 
multipole limit is reached for relatively low values of $k$, $k\sim 10$, and 
this result is insensitive to the size of the non-linear multipoles.  
Assuming the high multipole limit is a good measure of the dynamic 
aperture of an accelerator, then the dynamic aperture of an accelerator 
should be insensitive to the size of the higher multipoles, $k$ larger 
than 10 for RHIC.

\begin{figure}[tbh]
\centerline{\psfig{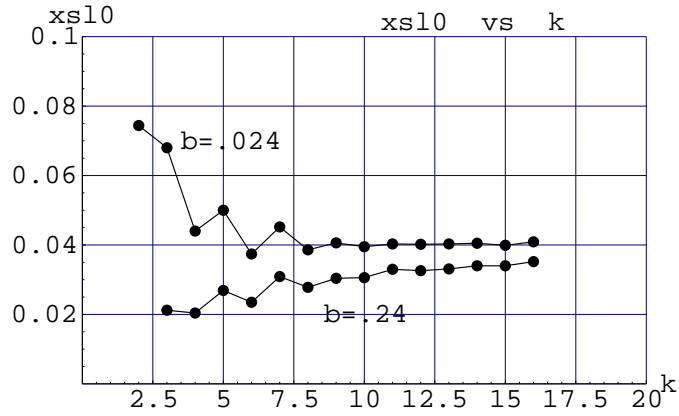}}
\caption{A plot showing how $x_{sl0}$ versus $k$ depends on the strength, 
$\mathbf{b}$, of the single multipole at $q_f$ for the simple one cell lattice, by 
comparing two cases where $\mathbf{b}$ differs by a factor of 10.  In the figure
xsl0 represents $x_{sl0}$.}
\label{fig:figtwo}
\end{figure}

Fig.~\ref{fig:figtwo} shows $x_{sl0}$ plotted against $k$ for the simple one 
cell lattice with a single $b_k$ at $q_f$.  The two plots shown are for 
$\mathbf{b}=0.024$ m and for $\mathbf{b}=0.24$ m.

One sees that the same high multipole limit is reached at about $k=10$ for 
$\mathrm{b}=.024$ and about $k=24$ for $\mathrm{b}=0.24$.

\begin{figure}[tbh]
\centerline{\psfig{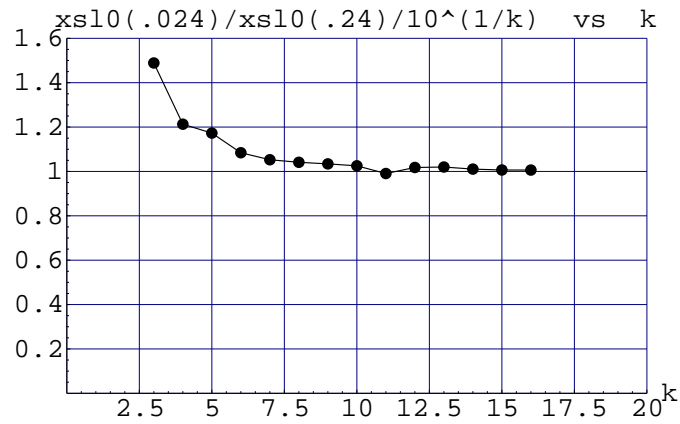}}
\caption{A plot showing that $x_{sl0}$ depends on the multipole strength 
like $x_{sl0}\sim \mathrm{b}^{(1/k)}$ for large $k$, by comparing two cases where 
$\mathbf{b}$ differs by a factor of 10.  In the figure, xsl0 represents $x_{sl0}$.}
\label{fig:figthree}
\end{figure}

Fig.~\ref{fig:figthree} shows that $x_{sl0}\sim \mathrm{b}^{(1/k)}$ by plotting 
$x_{sl0}(0.024)/x_{sl0}(0.24)/10^{(1/k)}$ against $k$. This ratio should 
approach 1 for large $k$ if $x_{sl0}\sim \mathrm{b}^{(1/k)}$.

In the above it was shown that $x_{sl0}\sim \mathbf{b}^{(1/k)}$ which shows the 
dependence of $x_{sl0}$ on $\mathrm{b}$ for a fixed $k$.  A more complete result 
which also shows the dependence on $k$ is
\begin{equation}
x_{sl0} \sim \mathrm{b}^{(1/k)}/ k
\label{eq:seventh}
\end{equation}

\bigskip

\noindent\underline{The dependence of the high multipole limit stability 
boundary on the choice}\\
\underline{of linear tunes, $\nu_x$ and $\nu_y$}

\medskip

Tracking studies done with the simple one cell lattice indicate that the 
stability boundary in the high multipole limit does not depend on the choice 
of linear tunes, $\nu_x$ and $\nu_y$. This leads to the suggestion, see 
section 8,  that the linear tunes  be chosen to avoid the resonances driven 
by the lower multipoles. The term lower multipoles is defined below.

\chapter{The high multipole limit in 4 dimensions}

In 2 dimensional phase space motion, the stability boundary that encloses 
the stable area in $x_0$, $p_{x0}$ in the high multipole limit is given by
\begin{equation}
\epsilon(x_0,p_{x0})=[R^2/\beta_x]_{\rm min}
\label{eq:eighth}
\end{equation}
One may ask what is the stability boundary for motion in 4 dimensional 
phase space in the high multipole limit.  To answer this, one has to 
consider the motion of a particle  moving in a lattice whose only 
nonlinear field in each element is that of a single multipole given by
\begin{equation}
B_y+iB_x= -B_0 \ \mathbf{b} ( (x+iy)/R)^k
\label{eq:ninth}
\end{equation}
where $R$ may depend on $s$, and so may $\mathbf{b}$  although not by very large 
factors. To find the high multipole limit stability boundary, one may use 
the argument given in section 2.  For very large $k$, and for $x^2+y^2$ not 
close to $R^2$, Eq.~\ref{eq:ninth} shows that $B_y\sim B_x\sim 0$. Thus for 
$x^2+y^2$ not close to $R^2$,
\begin{equation}
\epsilon_x(x,p_x)=\epsilon_{x0}, \qquad \epsilon_y(y,p_y)=\epsilon_{y0}
\label{eq:tenth}
\end{equation}
where $\epsilon_{x0}$, $\epsilon_{y0}$ are two constants and $\epsilon_x$ and 
$\epsilon_y$ are the linear emittance invariants. In addition for stable 
motion one has
\begin{equation}
( x^2+y^2) < R^2
\label{eq:eleven}
\end{equation}
Equation~\ref{eq:eleven} can be restated as, using Eq.~\ref{eq:tenth},
\begin{equation}
(\beta_x \ \epsilon_{x0}+\beta_y \  \epsilon_{y0}) < R^2
\label{eq:twelve}
\end{equation}
Eq.~\ref{eq:eleven} limits the range of $\epsilon_{x0}$, $\epsilon_{y0}$ for 
which the motion is stable. On the stability boundary , $\epsilon_x$ and 
$\epsilon_y$ are constant with the values of $\epsilon_{x0}$ and $\epsilon_{y0}$
respectively, and for each set of values of $\epsilon_{x0}$ and $\epsilon_{y0}$ 
that are on the stability boundary, $x^2+y^2$ must be less than or equal to 
$R^2$ for each element and at some point around the lattice $x^2+y^2$ must 
be equal to $R^2.$

\begin{figure}[tbh]
\centerline{\psfig{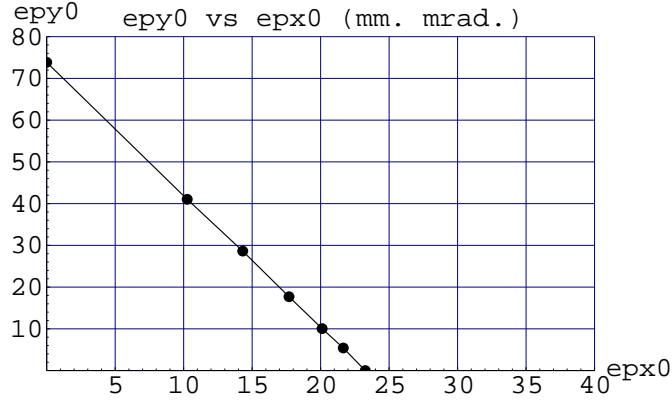}}
\caption{A plot showing the stability boundary in the hml for 4
dimensional motion, for the simple one cell lattice, as a plot of
$\epsilon_{y0}$ versus $\epsilon_{x0}$.  epy0, epx0 represent
$\epsilon_{y0}$, $\epsilon_{x0}$.}
\label{fig:figfour}
\end{figure}

Eqs.~\ref{eq:tenth} through \ref{eq:twelve} define the stability boundary 
in the high multipole limit. The stability boundary for the high multipole 
limit may be visualized as a curve in $\epsilon_{x0}$, $\epsilon_{y0}$ space 
as shown in Fig.~\ref{fig:figfour}. For motion in 4 dimensional phase space, 
there does not appear to be a simple solution  for the stability boundary 
in $\epsilon_{x0}$, $\epsilon_{y0}$ space.  This solution may depend on the 
particular form of $\beta_x(s)$, $\beta_y(s)$. There are, however, 3 points on 
the stability  boundary for which one can find simple results. These are the 3 
points for which 1. $\epsilon_{x0}=0$, 2. $\epsilon_{y0}=0$ and 3. 
$\epsilon_{x0}=\epsilon_{y0}$.

For the $\epsilon_{y0}=0$ point, Eqs.~\ref{eq:tenth} through \ref{eq:twelve} give
\begin{equation}
\epsilon_{x0}=[R^2/\beta_x]_{\rm min}, \qquad \epsilon_{y0}=0
\label{eq:thirteen}
\end{equation}

\noindent as the value of $\epsilon_{x0}$ on the stability boundary when 
$\epsilon_{y0}=0$. $[R^2/\beta_x]_{\rm min}$ is the minimum value of 
$R^2/\beta_x$  around the lattice. With this value of $\epsilon_{x0}$ and 
$\epsilon_y=0$,  one can show that $(x^2+y^2)$ is less than or equal to 
$R^2$ at every element in the lattice. Similarly for $\epsilon_{x0}=0$, one finds
\begin{equation}
\epsilon_{y0}=[R^2/\beta_y]_{\rm min}, \qquad \epsilon_{x0}=0
\label{eq:fourteen}
\end{equation}
When $\epsilon_{x0}=\epsilon_{y0}$, one finds
\begin{equation}
\epsilon_{x0}=\epsilon_{y0}= [R^2/(\beta_x+\beta_y)]_{\rm min}
\label{eq:fifteen}
\end{equation}

\noindent The three  points on the stability boundary given by 
Eqs.~\ref{eq:thirteen} through \ref{eq:fifteen} provide a fairly good picture 
of the stability boundary.  A fairly good approximation can be obtained by 
drawing two straight lines between the known three points.

For accelerators  which have insertion regions where $\beta_x$, $\beta_y$ 
have exceptionally large values which occur in magnets which have the same 
$R$ value, $R_{\rm ins}$,  then Eqs.~\ref{eq:thirteen} through \ref{eq:fifteen}
can written as
\begin{eqnarray}
\epsilon_{x0} & = & R_{\rm ins}^2/\beta_{x{\rm max}}, \qquad \epsilon_{y0}=0 
\nonumber \\
\epsilon_{y0} & = & R_{\rm ins}^2/\beta_{y{\rm max}}, \qquad \epsilon_{x0}=0 
\label{eq:sixteen}\\
\epsilon_{x0} & = & \epsilon_{y0}=R_{\rm ins}^2/(\beta_x+\beta_y)_{\rm max} 
\nonumber
%\label{eq:sixteen}
\end{eqnarray}
$\beta_{x{\rm max}}$ is the largest $\beta_x$ in the lattice, and $\beta_{y{\rm max}}$ 
and $(\beta_x+\beta_y)_{\rm max}$ have similar meanings.

Often, in tracking studies one does runs with initial values $\epsilon_{x0 }=
\epsilon_{y0}$ and $p_{x0}=0$, $p_{y0}=0$, increasing $x_0$ until the motion becomes 
unstable for a given number of periods. The corresponding value of $x_0$ may 
be labeled $x_{sl0}$. If one does tracking studies where all the $b_k$ present 
have the same $k$ value, then for large enough $k$, the basic rule high 
multipole limit in 4 dimensions states that $x_{sl0}$ will approach a 
non-zero value given by Eq.~\ref{eq:fifteen} as
\begin{equation}
x_{sl0}=\sqrt{(\beta_{x0}/R_0^2) [R^2/(\beta_x+\beta_y)]_{\rm min}}
\label{eq:seventeen}
\end{equation}
where $\beta_{x0}$ and $R_0$ are these parameters at the element where $x_{sl0}$ 
is measured. For the insertion case described by Eq.~\ref{eq:sixteen}, 
$x_{sl0}$ is given by
\begin{equation}
x_{sl0}=(R_0/R_{\rm ins}) \sqrt{\beta_{x0}/(\beta_x+\beta_y)_{\rm max}}
\label{eq:eighteen}
\end{equation}

In Fig.~\ref{fig:figfour}, $\epsilon_{y0}$ is plotted against $\epsilon_{x0}$ 
showing the stability boundary in the initial emmitance space of 
$\epsilon_{x0}$, $\epsilon_{y0}$. This curve was found by a tracking study 
using the simple one cell lattice with a single $b_k$ at $q_f$.  $k$ was 
chosen at the large value of $k=1\ 10^6$, so that the curve is a good 
approximation of the high multipole limit.  In this case the stability 
boundary in $\epsilon_{x0}$, $\epsilon_{y0}$ is almost a straight line. A 
simple approximation of the stability boundary is a straight line 
connecting the end points, the $\epsilon_{x0}=0$ point and the $\epsilon_{y0}=0$ 
point. According to Eqs.~\ref{eq:thirteen} through \ref{eq:fifteen} this 
straight line is given by
\begin{equation}
\epsilon_{x0}/[R^2/\beta_x]_{\rm min} + \epsilon_{y0}/[R^2/\beta_y]_{\rm min} = 1
\label{eq:ninteen}
\end{equation}
For the simple one cell lattice used here, where $b_k$ is not zero only at 
the $qf$ element, then $[R^2/\beta_x]_{\rm min}=R^2/\beta_{xqf}$ and 
$[R^2/\beta y]{\rm min}=R^2/\beta_{yqf}$, $R=0.04$ m and $\beta xqf=68.4497$, 
$\beta yqf= 21.6265$. One finds in this case that the three points on the  
stability boundary as given by Eqs.~\ref{eq:thirteen} through \ref{eq:fifteen}
agree with the tracking results with an error less than $5\ 10^{-3}$.

One may note that if $[R^2/\beta_x]_{\rm min}= [R^2/\beta_y]_{\rm min}$, 
which is true for RHIC because $\beta x=\beta y$ at the low beta crossing 
points and tends to be true for proton colliders, then Eq.~\ref{eq:ninteen} 
gives
\begin{equation}
\epsilon_{x0} + \epsilon_{y0} = {\rm constant}
\label{eq:twenty}
\end{equation}
on the stability boundary.  Tracking studies indicate that Eq.~\ref{eq:twenty} 
is roughly true for RHIC.  Assuming the high multipole limit provides a 
reasonable estimate of the dynamic aperture of the actual accelerator, then 
the above shows that the result that the total emittance, 
$\epsilon_{x0} + \epsilon_{y0}$, is roughly constant on the stability boundary 
is accidental in the sense that it depends on the properties of the beta 
functions at the crossing points. If these beta functions are not equal, 
then Eq.\ref{eq:ninteen} will replace Eq.~\ref{eq:twenty}.

\begin{figure}[tbh]
\centerline{\psfig{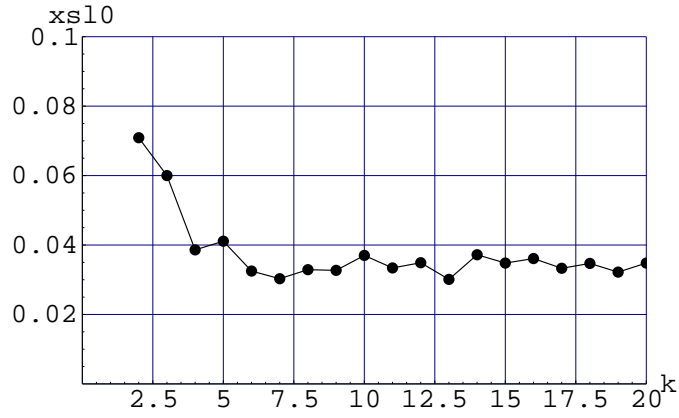}}
\caption{A plot of $x_{sl0}$ versus $k$ for the simple one cell lattice with a 
single multipole of order $k$ at $q_f$, for 4 dimensional motion when 
$p_{x0}=0 =p_{y0}$ and $\epsilon_{y0}=\epsilon x0$.  In the figure, 
xsl0 represents $x_{sl0}$ and is in meters.}
\label{fig:figfive}
\end{figure}

Figure~\ref{fig:figfive} illustrates the result given by 
Eq.~\ref{eq:seventeen} for $x_{sl0}$ for the case when $\epsilon_{x0}=
\epsilon_{y0}$, and $p_{x0}=p_{y0}=0$ in the high multipole limit. In fig.~\ref{fig:figfive} 
$x_{sl0}$ is plotted against the multipole order, $k$. The results were found 
in a tracking study using the simple one cell lattice with a single $b_k$ 
at $q_f$. For this lattice, with the parameters used, $R=0.04$ m, and at $q_f$, 
$\beta_x,\beta_y= 68.4497$, 21.6265 m. Eq.~\ref{eq:seventeen} then gives 
for $x_{sl0}$ at very large $k$, $x_{sl0}=.0349$ m.  Fig.~\ref{fig:figfive} shows 
that $x_{sl0}$ is approaching a value at large $k$ near 0.0349m. At $k=20$, 
$x_{sl0}=.0348$ m was found.

\chapter{High multipole limit and the dynamic aperture}

The high multipole limit gives a result for the stability boundary that 
encloses the stable area in $x_0$, $p_{x0}$.  The goal of this section is to 
show that the high multipole limit gives a reasonable approximation for 
the stability boundary when all the field multipoles are present and the 
lower multipoles have been corrected.  Conversely, the high multipole 
limit indicates how much may be gained by correcting the lower multipoles. 
The phrase lower multipoles will be more precisely defined below. Also one 
will see that when the nonlinear field multipoles are not too large, as is 
the case in RHIC, the high multipole limit provides a rough but useful 
estimate of the dynamic aperture.

\bigskip

\noindent\underline{Motion in 2 dimensional phase space}

\medskip

The above statements can be illustrated by the results of a tracking study 
in 2 dimensional phase space using the simple one cell lattice with non 
linear multipoles only at $q_f$.  To simulate the multipoles present in an 
accelerator, the point like nonlinear field at $q_f$  is given by
\begin{equation}
B_y=B_0\ mathrm{b} (x/R)^2 (1-(x/R)^{50}) / (1-x/R)
\label{eq:tone}
\end{equation}
Eq.~\ref{eq:tone} gives a nonlinear field which contains all multipoles from 
$k=2$ to about $k=50$, where all the multipoles decrease like $1/R^k$.  With 
the parameters chosen as $R=0.04$m, $\mathrm{b}=0.024$, this lattice resembles  
the RHIC accelerator without insertions.
\begin{figure}[tbh]
\centerline{\psfig{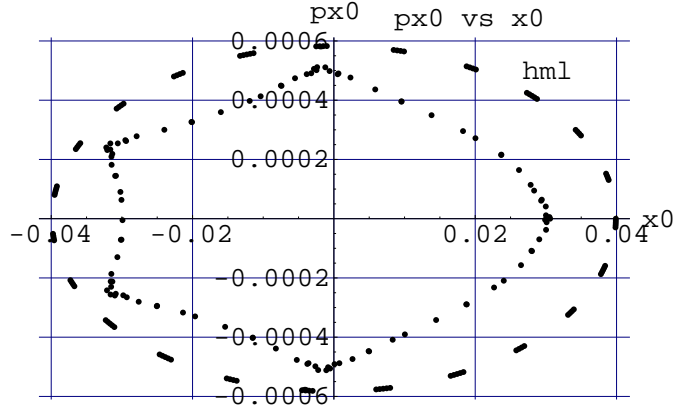}}
\caption{A plot of $p_{x0}$ versus $x_o$ showing the stability boundary for 2 
dimensional motion in the simple one cell lattice with all multipoles from 
$k=2$ to $k=50$ at $q_f$.  The stability boundary in the high multipole 
limit is also shown.  In the figure, px0,x0 represent $p_{x0}$, $p_0$.}
\label{fig:figsix}
\end{figure}

Fig.~\ref{fig:figsix} shows the stability boundary in $x_0$, $p_{x0}$ space as 
measured at $q_f$. Two boundaries are shown; one is the high multipole limit, 
and the other boundary, that is contained inside the hml boundary, is the 
stability boundary when all the multipoles are present as given by 
Eq.~\ref{eq:tone}.  According to the suggestions made at the beginning of 
this section, the difference between these two boundaries shows the loss in 
stable phase space due to the lower multipoles, and also how much phase 
space can be gained by correcting the lower multipoles.

\begin{figure}[tbh]
\centerline{\psfig{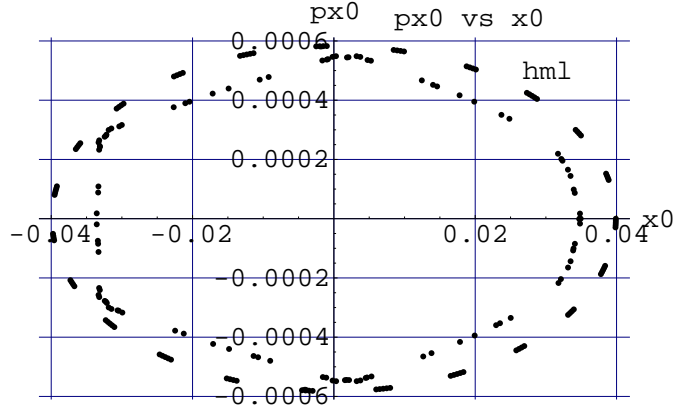}}
\caption{A plot of $p_{x0}$ versus $x_o$ showing the stability boundary for 2 
dimensional motion in the simple one cell lattice with multipoles from 
$k=10$ to $k=50$ at $q_f$. The lower multipoles, $k=2$ to $k=9$ have been 
corrected.  The stability boundary in the high multipole limit is also shown. 
 In the figure, px0,x0 represent $p_{x0}$, $x_0$.}
\label{fig:figseven}
\end{figure}

The term lower multipoles will be defined as follows.  If one looks at 
Fig.~\ref{fig:figone} and Fig.~\ref{fig:figfive} which plot $x_{sl0}$ vs $k$ 
for the two cases, $\epsilon_{y0}=0$ and $\epsilon_{y0}=\epsilon x0$, one sees 
that $x_{sl0}$ gets close to the value given by the high multipole limit at 
about $k=10$ for the simple one cell lattice.  This value of $k$ where 
$x_{sl0}$ gets close to the value given by the hml will be called 
$k_{\rm hml}$.  The lower multipoles are those multipoles for which $k$ is 
less than $k_{\rm hml}$.  Let us now correct the lower multipoles from $k=2$ 
to $k=9$ giving the plot shown in Fig.~\ref{fig:figseven}.

Fig.~\ref{fig:figseven} shows that by correcting $b_2$ to $b_9$, the stable phase 
space area has been increased so that it lies fairly close to the high 
multipole limit result, but still lies within the high multipole limit.  
If one corrects more multipoles past $b_9$, the stability boundary will 
increase approaching the result for the high multipole limit.  Results 
found using a RHIC lattice will be presented in section 6 which will also 
support the validity of the suggestions made in this section about the 
connection between the dynamic aperture and the high multipole limit. 
One may note that Fig.~\ref{fig:figsix} shows that the loss in stable 
phase space due to the non linear multipoles used is about a factor of 2. 
About the same factor will be found for a RHIC lattice.  Thus one can say 
that the hml provides a rough but useful estimate of dynamic aperture when 
the multipoles present are of the order of those expected in RHIC, 
overestimating the stable phase area in 2 dimensional phase space by 
about a factor of 2.

\bigskip

\noindent\underline{Motion in 4 dimensional phase space}

\medskip

The suggestions made at the beginning of this section can be illustrated by 
the results of a tracking study in 4 dimensional phase space using the 
simple one cell lattice with nonlinear multipoles only at $q_f$. To simulate 
the multipoles present in an accelerator, the point like nonlinear field at 
$q_f$ is given by
\begin{equation}
B_y+iB_x=B_0\  \mathrm{b} ((x+iy)/R)^2 (1-(x+iy)/R)^{50}) / (1-(x+iy)/R)
\label{eq:ttwo}
\end{equation}
Eq.~\ref{eq:ttwo} gives a nonlinear field which contains all multipoles 
from $k=2$ to about $k=50$, where all the multipoles decrease like $1/R^k$. 
With the parameters chosen as $R=0.04$m, $\mathrm{b}=0.024$, this lattice resembles 
the RHIC accelerator without insertions.

\begin{figure}[tbh]
\centerline{\psfig{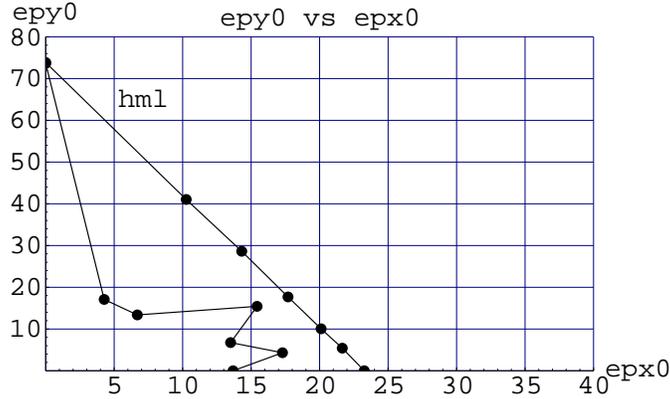}}
\caption{A plot of $\epsilon_{y0}$ versus $\epsilon_{x0}$ showing the stability 
boundary in $\epsilon_{x0}$, $\epsilon_{y0}$ space for 4 dimensional motion in 
the simple one cell lattice when all the multipoles from $k=2$ to $k=50$ 
are present at $q_f$. The stability boundary in the high multipole limit 
is also shown.  In the figure, epy0,epx0 represent $\epsilon_{y0}$, 
$\epsilon_{x0}$.}
\label{fig:figeight}
\end{figure}

Fig.~\ref{fig:figeight} shows the stability boundary in $\epsilon_{x0}$, 
$\epsilon_{y0}$ space as measured at $q_f$.  Two surfaces are shown; one is 
the high multipole limit, and other surface, that is contained within the 
high multipole limit surface, is the stability boundary when all the 
multipoles are present as given by Eq.~\ref{eq:ttwo}.  According to the 
suggestions made at the beginning of this section, the difference between 
these two boundaries shows the loss in stable phase space due to the lower 
multipoles, and also how much phase space can be gained by correcting the 
lower multipoles.

One may note that the high multipole limit stability boundary in 
$\epsilon_{x0}$, $\epsilon_{y0}$ space is a curve with zero thickness.  
However 
for the simple one cell lattice with all the multipoles present the 
$\epsilon_{y0}$ vs $\epsilon_{x0}$ stability curve has a certain smear or 
non-zero thickness.  This is because for a choice of $\epsilon_{x0}$, the 
$\epsilon_{y0}$ that lies on the stability boundary depends on the choice of 
$x_0$, $p_{x0}$, $y_0$, $p_{y0}$. The curve shown in Fig.~\ref{fig:figeight} 
was obtained with $p_{x0}=p_{y0}=0$.

\begin{figure}[tbh]
\centerline{\psfig{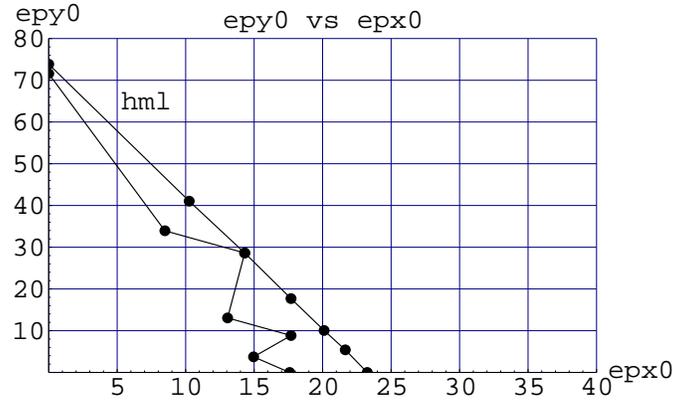}}
\caption{A plot of $\epsilon_{y0}$ versus $\epsilon_{x0}$ showing the 
stability boundary in $\epsilon_{x0}$, $\epsilon_{y0}$ space for 4 
dimensional motion in the simple one cell lattice when the multipoles 
from $k=10$ to $k=50$ are present at $q_f$.  The multipoles from $k=2$ 
to $k=9$ have been corrected. The stability boundary in the high multipole 
limit is also shown.  In the figure, epy0,epx0 represent $\epsilon_{y0}$, 
$\epsilon_{x0}$.}
\label{fig:fignine}
\end{figure}

Let us now correct the lower multipoles from $k=2$ to $k=9$ giving the plot 
shown in Fig.~\ref{fig:fignine}.   Fig.~\ref{fig:fignine} shows that by 
correcting $b_2$ to $b_9$, the stable phase space area has been increased so that 
it lies fairly close to the high multipole limit result, but still lies 
within the high multipole limit boundary.  If one corrects more multipoles 
past $b_9$, the stability boundary will increase approaching the result for the 
high multipole limit. Results found using a RHIC lattice will be presented 
in section 5 which will also support the validity of the suggestions made in 
this section about the connection between the dynamic aperture and the high 
multipole limit.  The results shown in Figs.~\ref{fig:figeight}, and 
~\ref{fig:fignine} also show how one can obtain misleading conclusions by 
looking at the results for just one direction in $\epsilon_{x0}$, 
$\epsilon_{yo}$ space like the $\epsilon_{x0}=\epsilon_{y0}$ direction. In this 
case the dynamic aperture actually became smaller for this direction when 
the $k=2$ to 9 multipoles were corrected.

\chapter{RHIC lattice results}

This section will illustrate the suggestions made in section 4 that the 
high multipole limit gives a reasonable estimate of the dynamic aperture 
when the lower multipoles are corrected  by giving the results of tracking 
studies done with an early version of the RHIC lattice \cite{key3}. This 
lattice has random non linear multipoles in each element from $k=2$ up to 
and including order $k=20$. Skew multipoles are also present.  The lattice 
has 6 insertions with 6 low beta crossing points at which $\beta_x= 6$ m 
and $\beta_{x{\rm max}}=236$m, $\beta_{y{\rm max}}=236$m, 
$(\beta_x+\beta_y)_{\rm max}= 309$ m. At $q_f$ in the normal cell, 
where $x$, $p_x$, $y$, $p_y$ are  observed, $\beta_x=56$ m, $\beta_y=8.72$m.

\bigskip

\noindent\underline{The high multipole limit in RHIC}

\medskip

In RHIC, the random multipoles in each element decrease roughly as $R^k$. 
Because the multipoles are chosen randomly corresponding to given rms 
values \cite{key3}, multipoles of different order or different $k$ values, 
have different strengths or different $\mathbf{b}$ values.  Elements with different 
$R$ values are also present. The elements that have the smallest values of 
$R^2/\beta_x$ or $R^2/\beta_y$, and which are the dominant elements, are in 
the insertions at the locations of $\beta_{xmx}$ and $\beta_{ymx}$. The RHIC 
lattice differs from the simple one cell lattice in the presence of the 
chromaticity correcting sextupoles. In RHIC, besides the random 
multipoles whose rms values decrease like $1/R^k$, there is also a set of 
multipoles, the chromaticity correcting sextupoles, which do not fit into 
the $1/R^k$ pattern of the random multipoles.  Because of this, it is 
necessary to change the definition of the high multipole limit for RHIC.  
As described above, the high multipole limit is found by doing a series of 
tracking studies in which all the elements of the lattice have one multipole 
present of the same order, $k$, and the particle motion for very large $k$ is 
the high multipole limit.  In the case of RHIC this procedure is changed in 
that in each tracking study the chromaticity correcting sextupoles are 
present as well  as the random multipoles of order $k$. With this change in 
the definition of the high multipole limit, it is again suggested that the 
high multipole limit  gives a reasonable estimate of the dynamic aperture 
when the lower multipoles are corrected.  This will be illustrated by the 
following tracking results found using a RHIC lattice. This procedure for 
defining the high multipole limit can be used for any lattice for which 
there is another nonlinear field present as well as a set of multipoles 
that decrease like $1/R^k$; for example, one systematic multipole may be 
exceptionally large.

This new definition of the high multipole limit changes the results found 
in sections 2 and 3 for the stability boundary in the high multipole limit. 
For very large $k$, and for $x^2+y^2$ smaller than $R^2$ in each element, 
the particle motion is that of a particle in the presence of the chromaticity 
correcting sextupoles. In addition for stable motion one has
\begin{equation}
( x^2+y^2) < R^2
\label{eq:tthree}
\end{equation}
To find the stability boundary, one needs to know what is the maximum value 
of $x^2+y^2$ for a given $x_0$, $p_{x0}$, $y_0$, $p_{y0}$. In section 3, this was 
given by the linear beta functions. In effect, one has to know what 
corresponds to the beta functions for the chromaticity correcting sextupoles 
for computing  the maximum value of $x^2+y^2$.  A semi-empirical solution 
of this problem will given below. One might notice one needs to answer this 
question only if one wants to have analytical results for points on the 
stability boundary in the high multipole limit like those given by 
Eqs.~\ref{eq:thirteen} through ~\ref{eq:fifteen}. One can always find the 
stability boundary in the high multipole limit with tracking studies without 
much difficulty.

\begin{figure}[tbh]
\centerline{\psfig{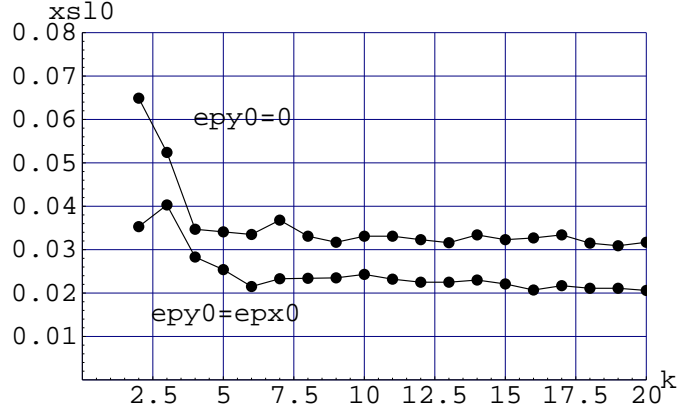}}
\caption{A plot of $x_{sl0}$ versus $k$ for RHIC.  Each element has a single 
multipole of the same order, $k$. For one curve $\epsilon_{y0}=0$ and 
$p_{x0}=p_{y0}=0$. For the other curve $\epsilon_{y0}=\epsilon_{x0}$ and $p_{x0}=p_{y0}=0$.   $x_{sl0}$ is computed for 500 turns.  In the figure, 
xsl0,epy0,epx0 represent $x_{sl0}$, $\epsilon_{y0}$, $\epsilon_{x0}$.}
\label{fig:figten}
\end{figure}

Fig.~\ref{fig:figten} shows $x_{sl0}$ plotted against $k$. In this study each 
element contains only one multipole of order $k$, and the multipoles in all 
the elements all have the same $k$ value and the chromaticity correcting 
sextupoles are also present. Two curves are shown. For one curve 
$\epsilon_{y0}=0$, and for the second curve, $\epsilon_{y0}=\epsilon_{x0}$.

$x_{sl0}$ is the largest $x_0$ that is stable for 500 turns when $p_{y0}=p_{x0}=0$. 
Fig.~\ref{fig:figten} is similar to Fig.~\ref{fig:figfive} found for the 
simple one cell lattice and shows that $x_{sl0}$ approaches a non zero limit 
as $k$ becomes large. Using the results found by tracking runs one can find 
a result for computing $x_{sl0}$ or $y_{sl0}$ in the high multipole limit in RHIC.  
For the case when the chromaticity correcting sextupoles are absent, $x_{sl0}$ 
and $y_{sl0}$ are given by Eq.~\ref{eq:sixteen} as
\begin{eqnarray}
x_{sl0} & = & (R_0/R_{\rm ins}) \sqrt{\beta_{x0}/\beta_{x {\rm max}}}, \quad 
\epsilon_{y0}=0 \nonumber \\
x_{sl0} & = & (R_0/R_{\rm ins}) \sqrt{\beta_{x0}/(\beta_x+\beta_y)_{\rm max}}, \quad
\epsilon_{y0 }=\epsilon_{x 0} \\
y_{sl0} & = & (R_0/R_{\rm ins}) \sqrt{\beta_{y0}/\beta_{y {\rm max}}}, \quad 
\epsilon_{x0}=0 \nonumber
\label{eq:tfour}
\end{eqnarray}
where $\beta_{x0}$, $\beta_{y0}$ and $R_0$ are these parameters at the 
element where $x_{sl0}$ is measured and for the insertion case described in 
section 3.  To obtain a result that may be valid when chromaticity correcting 
sextupoles are present, we will replace Eqs.~\ref{eq:tfour} by
\begin{eqnarray}
x_{sl0} & = & f_1 \ (R_0/R_{\rm ins}) \sqrt{\beta_{x0}/\beta_{x {\rm max}}}, \quad
\epsilon_{y0}=0 \nonumber \\
x_{sl0} & = & f_2 \ (R_0/R_{\rm ins}) \sqrt{\beta_{x0}/(\beta_x+\beta_y)_{\rm max}}, 
\quad \epsilon_{y0}=\epsilon_{x0} \label{eq:tfive}\\
y_{sl0} & = & f_3 \ (R_0/R_{\rm ins}) \sqrt{\beta_{y0}/\beta_{y{\rm max}}}, \quad
\epsilon_{x0}=0 \nonumber
%\label{eq:tfive}
\end{eqnarray}
In Eq.~\ref{eq:tfive}, $f_1$, $f_2$, $f_3$ can be found by using the results 
for $xsl0$ and $ysl0$ found with tracking studies done for RHIC. This gives 
the results
\begin{eqnarray}
x_{sl0} & = & (R_0/R_{\rm ins}) \sqrt{\beta_{x0}/\beta_{x {\rm max}}}, \quad
\epsilon_{y0}=0 \nonumber \\
x_{sl0} & = & 0.707 (R_0/R_{\rm ins}) \sqrt{\beta_{x0}/(\beta_x+\beta_y)_{\rm max}},
\quad \epsilon_{y0}=\epsilon_{x0} \label{eq:tsix}\\
y_{sl0} & = & (R_0/R_{\rm ins}) \sqrt{\beta_{y0}/\beta_{y{\rm max}}}, \quad
\epsilon_{x0}=0 \nonumber
%\label{eq:tsix}
\end{eqnarray}

Eqs.~\ref{eq:tsix} may be understood in the following way.  The chromaticity 
correcting sextupoles in RHIC do not greatly distort the particle motion. 
The stability surface in 2 dimensional phase space is a mildly distorted 
ellipse, as  will be seen below. The main effect of the chromaticity 
correcting sextupoles is due to the coupling  of the $x$ and $y$ motions, 
so that $x^2+y^2$ will grow from $x_0^2+y_0^2$ by a factor which is found to 
be close to 1.414. Thus in Eqs.~\ref{eq:tsix}, the $\epsilon_{y0}=0$ and the 
$\epsilon_{x0}=0$ results are unchanged from those found when the chromaticity 
correcting sextupoles are absent, while in the $\epsilon_{x0}=\epsilon_{y0}$ case,
the $x,y$ coupling changes the result for $xsl0$ by the factor 0.707. 
Although Eqs.~\ref{eq:tsix} were found using tracking results for RHIC, 
they may be used for other proton storage rings when the chromaticity 
correcting sextupoles play about the same role in distorting the particle 
motion.  Note that the factor of 0.707 in Eqs.~\ref{eq:tsix} is not based 
on analytical considerations, but was found though tracking studies with RHIC.

\begin{figure}[tbh]
\centerline{\psfig{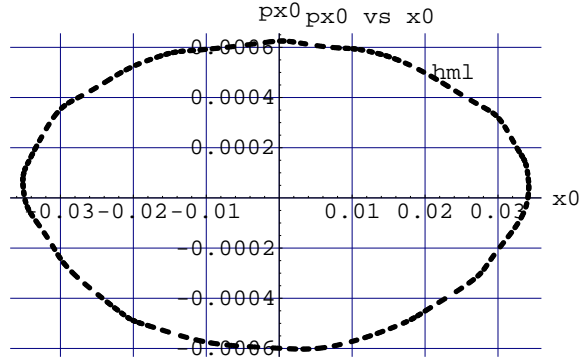}}
\caption{A plot showing the stability boundary for 500 turns in the high 
multipole limit for motion in 2 dimensional phase space for a RHIC lattice. 
 In the figure, px0,xo represent $p_{x0}$, $x_0$.}
\label{fig:figeleven}
\end{figure}

Fig.~\ref{fig:figeleven} shows the stability boundary in the high multipole 
limit for motion in 2 dimensional phase space, $\epsilon_{y0}=0$. Tracking 
runs of 500 turns were used. The stability boundary is almost elliptical, 
showing that the chromaticity correcting sextupoles do not distort the motion 
very much.

\begin{figure}[tbp]
\centerline{\psfig{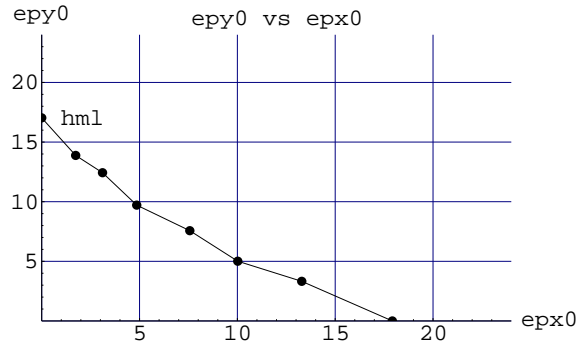}}
\caption{A plot showing the stability boundary in the high multipole limit 
for motion in 4 dimensional phase space for a RHIC lattice in 
$\epsilon_{y0}$ vs $\epsilon_{x0}$ space. Tracking runs of 500 turns were used 
with $p_{x0}=p_{y0}=0$ and with a single multipole with $k=20$ present in each 
element.  In the figure, epy0,epx0 represent $\epsilon_{x0}$, $\epsilon_{x0}$ 
which are in mm mrad.}
\label{fig:figtwelve}
\end{figure}

Fig.~\ref{fig:figtwelve} shows the stability boundary in the high multipole 
limit for motion in 4 dimensional phase space, in $\epsilon_{y0}$ vs 
$\epsilon_{x0}$ space. Tracking runs of 500 turns were used with $p_{x0}=
p_{y0}=0$ 
and with a single multipole with $k=20$ present in each element.  This 
surface has some thickness or smear which can be found by doing tracking 
runs with $p_{x0}$ and $p_{y0}$ not 0. This curve is almost a straight line, and 
using Eqs.~\ref{eq:tfive}, it is described very well by Eq.~\ref{eq:ninteen}
which in this case can be written as
\begin{equation}
\epsilon_{x0}/[R_{\rm ins}^2/\beta_{xmx}] + \epsilon_{y0}/[R_{\rm ins}^2/\beta_{ymx}] 
= 1
\label{eq:tseven}
\end{equation}

\bigskip

\noindent\underline{Motion in 2 dimensional phase space}

\medskip

\begin{figure}[bhpt]
\centerline{\psfig{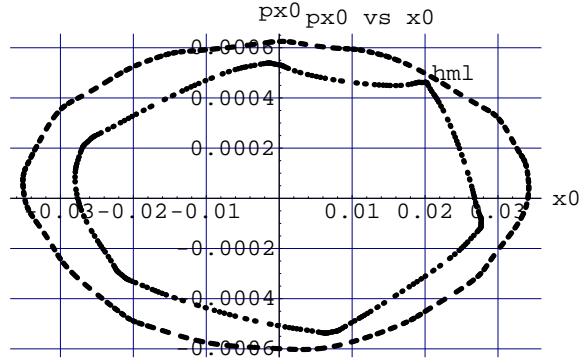}}
\caption{A plot showing the stability boundary for 500 turns for 2 
dimensional motion in RHIC with all the multipoles from $k=2$ to $k=20$ 
present. The stability boundary in the high multipole limit is also shown.  In the 
figure, px0,x0 represent $p_{x0}$, $x_0$.}
\label{fig:figthirteen}
\end{figure}

Tracking studies were done with a RHIC lattice to find the stability 
boundary for motion in 2 dimensional phase space, $y_0=p_{y0}=0$. The results 
are shown in Fig.~\ref{fig:figthirteen}. Two curves are shown. The outer 
curve is the stability boundary in the high multipole limit as was shown in 
Fig.~\ref{fig:figeleven}. The inner curve is the path in $x,p_x$  for the 
last $x_0$ that was stable for 500 turns as one increased $x_0$ with $p_{x0}=0$ 
and all the multipoles from $k=2$ to $k=20$ are present. According to the 
suggestion being made here about the significance of the  high multipole 
limit stability boundary, one would say that the lower multipoles have 
reduced the stable phase space by about 36\%.  This loss in phase space 
can be recovered by correcting the lower multipoles, $k$ less than about 
10.  Again, it is suggested here that the high multipole limit stability 
boundary indicates the stable phase space when the lower multipoles are 
corrected and it indicates the loss in phase space due to the lower 
multipoles.

Fig.~\ref{fig:figfourteen} is similar to Fig.~\ref{fig:figthirteen} except 
that the multipoles from $k=2$ to $k=10$ have been omitted.  One sees that 
as the lower multipoles are corrected, the stability boundary approaches that 
of the high multipole limit.  The loss in phase space has now been reduced to 
about 10\%.  Particle motions that came even closer to the hml and appeared 
to be stable for 500 turns were seen and were rejected because of a rather 
large smear and scatter.
\begin{figure}[bhp]
\centerline{\psfig{file=hmlFig5_5,width=3in}}
\caption{A plot showing the stability boundary for 500 turns for 2 
dimensional motion in RHIC with the multipoles from $k=11$ to $k=20$ present. 
The multipoles for $k=2$ to $k=10$ have been corrected. The stability 
boundary in the high multipole limit is also shown.  In the figure, px0,x0 
represent $p_{x0}$, $x_0$.}
\label{fig:figfourteen}
\end{figure}

\bigskip

\noindent\underline{Motion in 4 dimensional phase space}

\medskip

\begin{figure}[tbh]
\centerline{\psfig{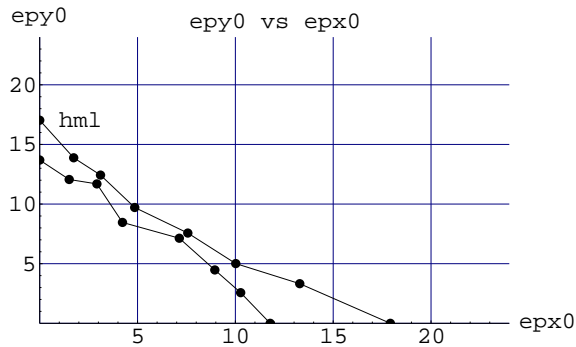}}
\caption{A plot showing the stability boundary for 500 turns for 4 
dimensional motion for a RHIC lattice with all the multipoles from 
$k=2$ to $k=20$ present. The stability boundary in the high multipole 
limit is also shown.  In the figure, epy0,epx0 represent $\epsilon_{y0}$, 
$\epsilon_{x0}$ are in mm-mrad.}
\label{fig:figfifteen}
\end{figure}

Fig.~\ref{fig:figfifteen} shows the results of tracking studies done with a 
RHIC lattice to find the stability boundary for motion in 4 dimensional  
phase space. The stability boundary is shown by plotting $\epsilon_{y0}$ 
versus $\epsilon_{x0}$ for the case where $p_{x0}=p_{y0}=0$. Two curves are shown. 
The outer boundary is the stability boundary in the high multipole limit. 
The inner boundary is the stability boundary when all the multipoles from 
$k=2$ to $k=20$ are present. The hml boundary was found by having only the 
$k=20$ multipole present. Fig.~\ref{fig:figfifteen} shows a loss in 4 
dimensional phase space of about 40\% due to the presence of the lower 
multipoles.

Fig.~\ref{fig:figsixteen} shows the result when the multipoles from 
$k=2$ to $k=10$ are corrected. One sees that as one corrects the lower 
multipoles, the stability boundary approaches the high multipole limit 
stability boundary, and the loss in 4 dimensional phase space is reduced 
to about 10\%.
\begin{figure}[tbh]
\centerline{\psfig{file=hmlFig5_7,width=3in}}
\caption{A plot showing the stability boundary for 500 turns for 4 
dimensional motion for a RHIC lattice with  the multipoles from $k=11$ to 
$k=20$ present.  The multipoles for $k=2$ to $k=10$ have been corrected. 
The stability boundary in the high multipole limit is also shown.  In the figure 
epy0,epx0 represent $\epsilon_{y0}$, $\epsilon_{x0}$ and are in mm-mrad.}
\label{fig:figsixteen}
\end{figure}

\chapter{Definition of stability}

In order to establish the properties of the high multipole limit, it is 
convenient to have a definition of stable motion which allows the stability 
boundary to be determined precisely. In considering the motion of a particle 
in an accelerator, one might consider the particle motion for a certain number
of periods to be stable if the particle motion stays within certain bounds, 
like those given by the vacuum tank, to be acceptable for the operation of 
the accelerator. Such a definition of stable motion, which allows a particular 
amount of acceptable growth in the particle motion, is not convenient for 
establishing the properties of the high multipole limit.  A definition of 
stable motion is given below which will precisely determine whether  a 
particle motion for a given number of periods is  stable. This definition 
may seem  artificial. However, a good deal of numerical tracking shows that 
the stability limits found  using this definition are usually  close to the 
stability limits that would be acceptable for an accelerator.

Consider the motion of a particle in a coordinate system which is based on a 
reference orbit where the independent coordinate is taken to be $s$, the 
distance along the reference orbit. The position of the particle is then 
described by $x$, $y$, and $s$, where $x$, $y$ are the coordinates along 
two directions perpendicular to the reference  orbit. The components of the 
momentum are then given by $p_x$, $p_y$ and $p_s$.  If the energy of the 
particle is assumed to remain constant during the tracking, then $p_s$ can be 
computed from $p_s = (p^2-p_x^2-p_y^2)^{0.5}$, $p$ being the total momentum of 
the particle. The motion of the particle over a given number of periods will 
be said to be unstable if during the tracking of the particle over the given 
number of periods, $p_s$ becomes imaginary or $p_x^2+p_y^2$ becomes larger than 
$p^2$.  $p_s$ becoming imaginary means that the formulation of the equations 
of motion based on the given reference orbit has broken down because $p_s$ 
has changed sign, and the particle has reversed its direction along the 
reference orbit so that $s$ is decreasing with time.

The above definition of unstable motion over a given number of periods may 
seem artificial.  However it has the advantage that the stability of a 
particular particle motion over a given number of periods can be precisely 
determined.  Much experience with tracking also indicates that it is a useful 
definition and usually gives results that are close to the stability limits 
that would be acceptable for an accelerator.  This definition of stability 
is convenient for establishing the above results for the high multipole 
limit.  This precise definition of stability allows the stability limit to 
be calculated with great accuracy, and the tracking searches for the 
stability boundary can be automated.  In order to use this definition of 
stability, one has to use the exact equations of motion. If one uses the 
approximations often used for large accelerators, where the radical 
$(p^2-p_x^2-p_y^2)^{0.5}$ is expanded out assuming that $p_x/p$ and $p_y/p$ 
are much smaller than one, one will obtain invalid results as the expansion 
is not valid when $p^2-p_x^2-p_y^2$ is near zero and the radical is about to 
become imaginary.

The definition of stability being proposed here has the following advantages:
\begin{enumerate}
\item
It avoids having to decide whether a particular particle motion is unstable 
when some growth occurs and it is not obvious whether the growth is 
acceptable or not.
\item
With this definition of stability tracking searches for points on the 
stability boundary can be automated as it provides a simple test for stability.
\end{enumerate}
 The results found in this paper do not depend on the choice
of this definition of stability. The same results woud be found with any other
reasonable definition of stability.

\chapter{Transfer functions for lattice elements}

In doing the tracking studies, one needs to know the transfer functions for 
each element of the lattice. The transfer functions allow one to compute the 
final coordinates of the particle from the initial coordinates for each 
element. As was indicated in section 6, the transfer functions have to 
satisfy the exact equations of motion in order to use the definition of 
stability given in section 6.  This can be accomplished by using the 
procedure \cite{key4} of replacing a magnet with point magnets at the ends 
of the magnet separated by a drift space.  By breaking the magnet up into 
pieces, one can approach the exact solution of the equations of motion by 
making the pieces smaller. One change in this procedure
 will be used here, which is that the reference orbit used 
will be made up of a series of smoothly joining straight lines and circular 
arcs \cite{key5}. The transfer functions are then given in Ref.~\cite{key5}. 
The circular arcs of the reference orbit are located at the dipoles in the 
lattice, and each arc has the curvature $\rho$ which depends on the strength 
of the dipole.  $1/\rho=0$ at the quadrupoles and drift spaces.

It is assumed that each magnet is broken up into a number of pieces. A 
magnet piece going from $s=s_1$ to $s=s_2$ and of length $h=s_2-s_1$ is replaced 
by point magnets at the ends separated by a drift space of length $h$. In 
the following, $q_x=p_x/p$, $q_y=p_y/p$, $q_s=(1-q_x^2-q_y^2)^{0.5}$.

\bigskip

\noindent A.\ \ \underline{Transfer functions for point magnets}

\medskip

The transfer functions for a point magnet located at $s=s_1$ is
\begin{eqnarray}
x_2 & = & x_1, \ \ y_2=y_1 , \nonumber \\
q_{x2} & = & q_{x1} + {1\over B\rho} {h\over 2} (1+x_1/\rho)
{\sin\theta\over\theta} B_y (x_1,y_1) , \label{eq:teight}\\
q_{y2} & = & q_{y1} - {1\over B\rho} {h\over 2} (1+x_1/\rho)
{\sin\theta\over\theta} B_x (x_1,y_1) , \nonumber
%\label{eq:teight}
\end{eqnarray}
$h$ is the length of the magnet piece, $\theta=h/\rho$. The field 
components $B_y$ and $B_x$ are assumed to depend only on $x$, $y$ and 
do not change along the magnet, and that $B_s=0$.

\bigskip

\noindent B.\ \ \underline{Transfer functions for drift spaces}

\medskip

For a region along $s$ in the lattice where $1/\rho=0$ for the reference orbit
\begin{eqnarray}
q_{x2} & = & q_{x1}, \ \ x_2 = x_1+q_{x1} L_{12} , \nonumber \\
q_{y2} & = & q_{y1}, \ \ y_2 = y_1+q_{y1} L_{12} , \nonumber \\
L_{12} & = & (s_2-s_1)/q_{s1} \label{eq:tnine}\\
q_s & = & (1-q_x^2-q_y^2)/^{1/2} , \nonumber
%\label{eq:tnine}
\end{eqnarray}
$L_{12}$ is the path length between $s_1$ and $s_2$.

For  a region  where $1/\rho$  is not zero,
\begin{eqnarray}
q_{x2} & = & q_{x1}\cos\theta + q_{s1}\sin\theta , \nonumber \\
q_{s2} & = & -q_{x1}\sin\theta + q_{s1}\cos\theta , \nonumber \\
\theta & = & (s_2-s_1)/\rho , \nonumber \\
x_2 & = & x_1+(1+x_1/\rho)2\rho\sin(\theta/2) \label{eq:thone}\\
& & \qquad \mbox{} \times {q_{x1}\cos\theta/2+q_{s1}\sin\theta/2\over
-q_{x1}\sin\theta + q_{s1}\cos\theta} , \nonumber \\
L_{12} & = & (1+x_1/\rho)\rho\sin (\theta) /q_{s2} , \nonumber \\
q_{y2} & = & q_{y1}, \ \ y_2=y_1+q_{y1} L_{12} \nonumber
%\label{eq:thone}
\end{eqnarray}

\bigskip

\noindent C.\ \ \underline{Transfer functions for the simple one cell lattice}

\medskip

This lattice has only point quadrupoles and drift spaces. For the transfer 
functions of the point quadrupoles one can use Eqs.~\ref{eq:teight}, 
replacing $(h/2) B_y$ and $(h/2) B_x$  by the integrated fields of the point 
magnet. For the drift spaces one can use Eqs.~\ref{eq:tnine}. The 
initial parameters that were used for the simple one cell lattice are the 
following:

\indent\indent quadrupole integrated strength = 436.647 KG

\indent\indent drift space length = 20m

\indent\indent multipole field, $b_k=\mathrm{b}\  /\  R^k$, $\mathrm{b}=0.024$, \ $R=0.04$m

\indent\indent $B\rho = 8400$ KG. m

\indent\indent $B_0=35$ KG

\chapter{Longterm effects and the high multipole limit}

The stability boundary in the high multipole limit for 2 dimensional phase 
space does not appear to depend on nprd, the number of periods the particle 
is tracked.  However many tracking studies have indicated that the stability 
boundary shrinks slowly the longer the particle is tracked.  If one accepts 
the statement that the stability boundary in the high multipole limit  is the 
boundary that is approached when the lower multipoles are corrected,  then 
one can remove the apparent contradiction by the suggestion that the 
shrinking of the stability boundary, when nprd is increased, is due to 
the presence of the lower multipoles, and this effect can be reduced by 
correcting the lower multipoles.

The following tracking study done with the simple one cell lattice supports 
the previous statements. If one considers $x_{sl0}$, the largest $x0$ that 
is stable for a given number of periods when $p_{x0}=0$, then one finds 
that $x_{sl0}$ decreases as nprd is increased. Using nprd $= 10^2$ and 
nprd $= 10^4$,  one finds the $dx_0/x_0$, the fractional decrease in $x_{sl0}$ 
for  these two values of nprd is $dx_0/x_0=0.033$ when all the multipoles are 
present.  If one corrects  some of the lower multipoles by omitting the 
multipoles for k=2 to k=9, then one finds that $dx_0/x_0$ is decreased by a 
factor of 6 to $dx_0/x_0=0.005$.

\bigskip

\noindent\underline{Avoiding resonances of order 10 or higher}

\medskip

It is sometimes suggested that in choosing the operating point for 
superconducting proton storage rings, one should avoid resonances of order 
10 or higher.  A basis for this rule is provided by the high multipole 
limit. The range of the lower multipoles that reduce the dynamic aperture 
below that given by the high multipole limit is given by the parameter 
$k_{hml}$ defined in section 4.  For RHIC,  $k_{hml}$ is about $k_{hml}=10$. 
Since the important multipoles in affecting the dynamic aperture are the 10 
lowest multipoles, it would seem desirable to avoid resonances up to 10 or 
higher  which are the resonances driven by the 10 lowest multipoles in lowest 
order. If one would increase the strength of the non-linear multipoles by a 
factor  of 10, thus raising $k_{hml}$ to about $k_{hml}= 20$,  the above 
argument would suggest that one should avoid resonances up to order 20 or 
higher.

\end{document}